\documentclass[aps,prx,reprint,twocolumn,superscriptaddress,floatfix,nofootinbib,longbibliography,preprintnumbers]{revtex4-2}
\usepackage{graphicx,amsmath,amsfonts,amssymb,amsthm,xr}
\usepackage{epsfig,amsmath,amssymb,color,dsfont,upgreek,physics}
\usepackage{mathrsfs}
\usepackage{mathtools}
\usepackage{bbold}
\usepackage{comment}
\usepackage{float}
\usepackage{comment}
\usepackage{bm}

\usepackage[bookmarks=true,colorlinks,linkcolor=OrangeRed,urlcolor=NavyBlue,citecolor=RoyalBlue]{hyperref}
\usepackage[dvipsnames]{xcolor}
\usepackage{tikz}
\usetikzlibrary{matrix,fit,positioning}
\usepackage{nicematrix}
    
\begin{document}

\title{\textbf{Observation of quantum-field-theory dynamics on a spin-phonon quantum computer}
}

\author{Anton T. Than${}^*$}
\affiliation{Joint Quantum Institute, University of Maryland, College Park, MD 20742, USA}
\affiliation{Department of Physics, University of Maryland, College Park, MD 20742, USA}
\affiliation{National Quantum Laboratory (QLab), University of Maryland, College Park, MD 20740, USA}

\author{Saurabh V. Kadam${}^{\dagger}$}
\affiliation{
InQubator for Quantum Simulation (IQuS), Department of Physics, University of Washington, Seattle, WA 98195, USA}

\author{Vinay Vikramaditya${}^{\dagger}$}
\affiliation{Department of Physics, University of Maryland, College Park, MD 20742, USA}
\affiliation{National Quantum Laboratory (QLab), University of Maryland, College Park, MD 20740, USA}
\affiliation{Joint Center for Quantum Information and Computer Science, College Park, MD 20742, USA}

\author{Nhung H. Nguyen}
\affiliation{Joint Quantum Institute, University of Maryland, College Park, MD 20742, USA}
\affiliation{Department of Physics, University of Maryland, College Park, MD 20742, USA}

\author{Xingxin Liu}
\affiliation{Joint Quantum Institute, University of Maryland, College Park, MD 20742, USA}
\affiliation{Department of Physics, University of Maryland, College Park, MD 20742, USA}
\affiliation{National Quantum Laboratory (QLab), University of Maryland, College Park, MD 20740, USA}

\author{Zohreh Davoudi${}^{\ddagger}$}
\affiliation{Department of Physics, University of Maryland, College Park, MD 20742, USA}
\affiliation{National Quantum Laboratory (QLab), University of Maryland, College Park, MD 20740, USA}
\affiliation{Joint Center for Quantum Information and Computer Science, College Park, MD 20742, USA}

\author{Alaina M. Green${}^{\ddagger}$}
\affiliation{Joint Quantum Institute, University of Maryland, College Park, MD 20742, USA}
\affiliation{Department of Physics, University of Maryland, College Park, MD 20742, USA}
\affiliation{National Quantum Laboratory (QLab), University of Maryland, College Park, MD 20740, USA}

\author{Norbert M. Linke${}^{\ddagger}$}
\affiliation{Department of Physics, University of Maryland, College Park, MD 20742, USA}
\affiliation{National Quantum Laboratory (QLab), University of Maryland, College Park, MD 20740, USA}
\affiliation{Duke Quantum Center and Department of Physics, Duke University, Durham, NC 27701, USA}

\date{\today}

\begingroup
\renewcommand\thefootnote{*}
\footnotetext{Corresponding author; athan42@umd.edu.}
\endgroup

\begingroup
\renewcommand\thefootnote{$\dagger$}
\footnotetext{These authors contributed equally to this work.}
\endgroup

\begingroup
\renewcommand\thefootnote{$\ddagger$}
\footnotetext{These authors jointly supervised this work.}
\endgroup

\preprint{UMD-PP-025-05, IQuS@UW-21-107}

\begin{abstract}
Simulating out-of-equilibrium dynamics of quantum field theories in nature is challenging with classical methods, but is a promising application for quantum computers. Unfortunately, simulating interacting bosonic fields involves a high boson-to-qubit encoding overhead. Furthermore, when mapping to qubits, the infinite-dimensional Hilbert space of bosons is necessarily truncated, with truncation errors that grow with energy and time. A qubit-based quantum computer, augmented with an active bosonic register, and with qubit, bosonic, and mixed qubit-boson quantum gates, offers a more powerful platform for simulating bosonic theories. We demonstrate this capability experimentally in a hybrid analog-digital trapped-ion quantum computer, where qubits are encoded in the internal states of the ions, and the bosons in the ions' motional 
states. Specifically, we simulate nonequilibrium dynamics of a (1+1)-dimensional Yukawa model, a simplified model of interacting nucleons and pions, and measure fermion- and boson-occupation-state probabilities. These dynamics populate high bosonic-field excitations starting from an empty state, and the experimental results capture well such high-occupation states. This simulation approaches the regime where classical methods become challenging, bypasses the need for a large qubit overhead, and removes truncation errors. Our results, therefore, open the way to achieving demonstrable quantum advantage in qubit-boson quantum computing.
\end{abstract}

\maketitle

\section{\label{sec:intro}Introduction}
Bosonic degrees of freedom are present in a wide range of physical systems. For example, scalar field theories describe a plethora of physical phenomena, from magnetization~\cite{fradkin2023field} to hadronic interactions~\cite{leutwyler1994foundations,weinberg1979phenomenological} to cosmic inflation~\cite{guth2023inflationary,urena2016scalar}. Additionally, gauge field theories are the backbone of the Standard Model of particle physics, and the gauge bosons are the carriers of fundamental forces in nature~\cite{aitchison2012gauge}. Simulating out-of-equilibrium dynamics of interacting bosonic (and bosonic-fermionic) quantum field theories has far-reaching applications in early-universe and collider physics~\cite{bauer2023quantum,bauer2024quantum,halimeh2025quantum}. These simulations are currently out of reach of classical computing methods, but may become tractable with quantum simulations.

Quantum simulation of bosonic field theories requires tackling the challenge of the bosons' infinite-dimensional Hilbert space. In a digital quantum-simulation setup, the bosonic degrees of freedom are mapped to a finite number of qubits (or qudits), using a truncation scheme and its associated cutoff. The errors in a physical observable due to this truncation decrease with increasing cutoff~\cite{klco2019digitization,gupta2025euclidean,gupta2025euclidean}. In real-time processes with abundant energy and entanglement generation, an increasingly larger part of the bosons' Hilbert space is accessed during system's evolution, requiring increasingly larger cutoffs to retain a fixed accuracy~\cite{jordan2011quantum,tong2022provably,csahinouglu2021hamiltonian,hanada2023estimating}. Furthermore, bosonic operations often amount to complex digital quantum circuits~\cite{nielsen2010quantum,bhaskar2015quantum,cuccaro2004new,gidney2018halving}. These features have led to impractically large quantum-computing cost estimates for simulating (gauge-)field-theory dynamics~\cite{kan2021lattice,davoudi2023general,rhodes2024exponential,davoudi2025tasi,jordan2011quantum,watson2023quantum}. 

Analog quantum simulation offers an alternative approach~\cite{georgescu2014quantum,altman2021quantum}. When an analog quantum simulator hosts well-controlled, naturally occurring bosonic degrees of freedom, the simulator's Hamiltonian can be engineered to mimic that of the target bosonic theory, without the need for Hilbert-space truncation. Examples of analog quantum simulators with bosonic excitations are trapped ions~\cite{monroe2021programmable}, ultracold bosonic gasses~\cite{bloch2012quantum}, circuit QED~\cite{blais2021circuit}, and photonic platforms~\cite{wang2020integrated}. However, the intrinsic interactions and the level of quantum control may restrict the types of bosonic Hamiltonians that can be engineered on a given hardware architecture.

One can instead combine the flexibility offered by the digital approach with the efficiency of the analog approach in a hybrid scheme. Here, a discrete set of operations can be designed by leveraging the simulator's intrinsic degrees of freedom and interactions. This scheme allows for arbitrary Hamiltonian dynamics to be simulated, and enables nontrivial state preparation and observable measurements. Bosonic and spin-boson quantum computations have gained considerable attention in recent years, resulting in proposals for simulating physical systems~\cite{mezzacapo2012digital,casanova2012quantum,lamata2014efficient,bermudez2017quantum,davoudi2021toward,bazavan2024synthetic,martinez2024thermal,crane2024hybrid}, and the development of related universal, fault-tolerant frameworks~\cite{michael2016new,noh2020fault,crane2024hybrid,liu2024hybrid, matsos2025universal}. To bridge the theory and practice, robust experimental feasibility demonstrations are, however, critically needed. This work constitutes an experimental demonstration of a hybrid analog-digital quantum simulation of the Yukawa model, i.e., a fermionic-bosonic quantum field theory, in a well-suited platform, trapped ions.

A trapped ion quantum simulator hosts both qubit and bosonic degrees of freedom. The qubits are encoded in two electronic states of the ions while the bosonic modes are enocded in the ions' shared harmonic motional modes. Excitations of the motion, known as phonons, serve two purposes in a hybrid scheme.
The phonons directly couple to the qubits, or can be made to self-interact, via ion-laser interactions~\cite{wineland1998experimental,katz2023demonstration,matsos2025universal,buazuavan2024squeezing,hou2024coherent,Porras2004,davoudi2021toward}, hence undergo nontrivial dynamics. They also serve as a mediator to entangle the qubits~\cite{molmer1999multiparticle,cirac1995quantum}. 
These features, along with experimental progress in harnessing phonons in quantum simulation~\cite{haze2012observation,toyoda2015hong,zhang2016fermion,debnath2018observation,ohira2019phonon,fluhmann2019encoding,tamura2020quantum,chen2021quantum,alderete2021experimental,katz2023demonstration,sun2025quantum,macdonell2023predicting}, motivate the use of trapped ions as an ideal platform for spin-boson quantum computing. As examples, motional modes can serve as a reservoir~\cite{sun2025quantum} to prepare a thermal state of the qubits in a controlled manner~\cite{than2024phase}. They can also be used in adibatic preparation of the ground state of bosonic-fermionic theories~\cite{shen2018quantum,whitlow2023quantum,valahu2023direct}.

Here, we aim to treat phonons as dynamical degrees of freedom, and track their coherent dynamics through the evolution, as is required for bosonic-fermionic quantum-field-theory simulations. 
Concretely, we take advantage of the infinite-dimensional Hilbert space of the motional modes to simulate dynamics of scalar fields coupled to fermions~\cite{davoudi2021toward}. Our model is the well-known Yukawa model~\cite{weinberg1967model} that was originally formulated to describe the interactions between nucleons (fermions) mediated by pions (bosons), to model the strong nuclear force~\cite{yukawa1935interaction}. Moreover, in the Standard Model, quarks and charged leptons (fermions) acquire their mass via their Yukawa coupling to the Higgs field (boson)~\cite{higgs1964broken}. We focus on a Yukawa model in (1+1)D, where scalar fields are coupled to one flavor of fermions on a spatial lattice. We encode instances of this model in our spin-phonon quantum computer, and track non-trivial dynamics of both fermions and bosons after a quench of the Hamiltonian. Our results agree well with theoretical predictions, even when highly occupied bosonic states are created in the out-of-equilibrium dynamics. This work, therefore, demonstrates the value of efficient boson encoding in dynamical simulations, and represents a promising step toward the ultimate goal of leveraging spin-boson quantum computers to simulate complex physical systems.

\begin{figure*}[t!] 
    \centering
    \includegraphics[width=0.85\textwidth]{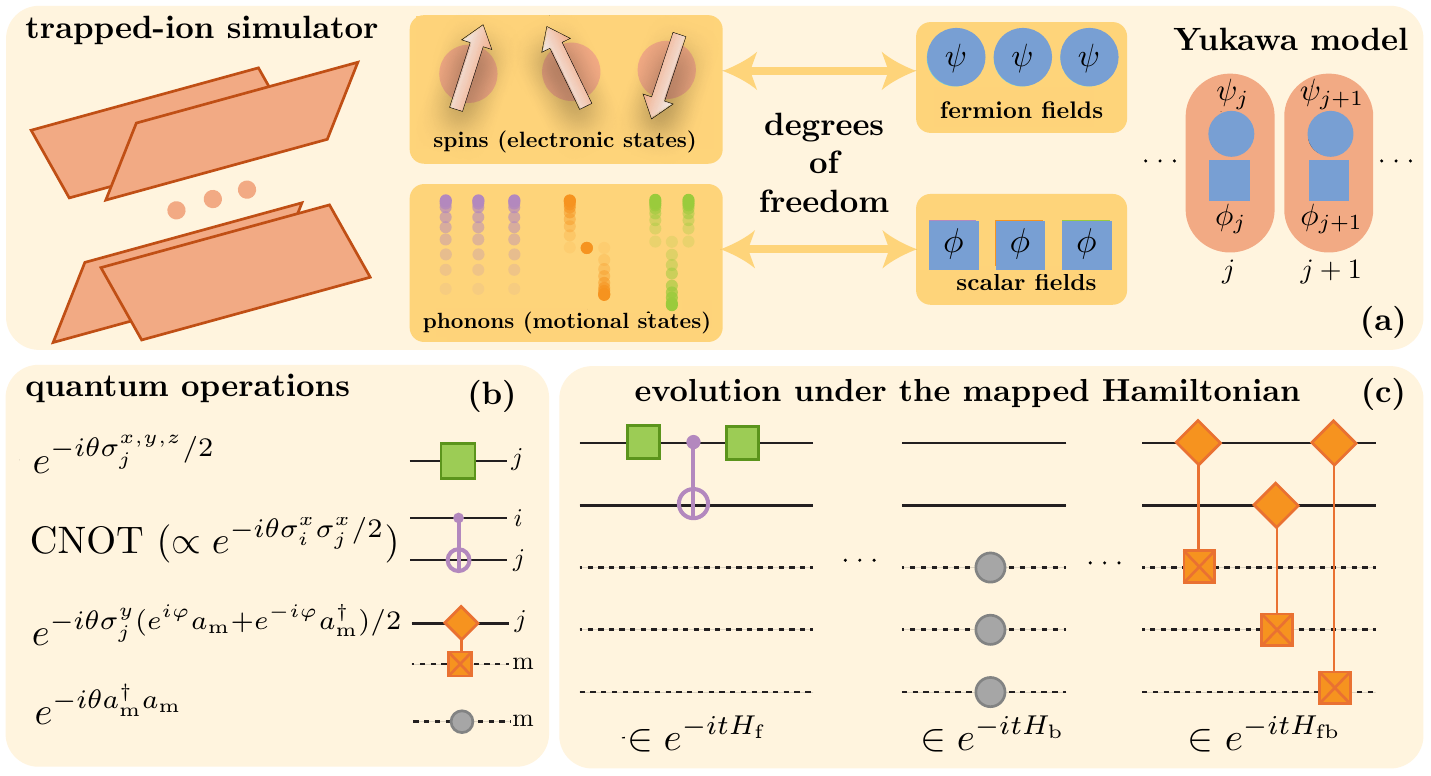} 
    \caption{\emph{Spin-phonon quantum-simulation setup}. (a) The degrees of freedom in a bosonic-fermionic quantum field theory are mapped to those in a hybrid analog-digital quantum computer based on trapped ions. Specifically, the fermion field, $(\psi_j)$, on site $j$ of a 1D lattice is mapped onto the electronic states of a corresponding ion in the quantum computer (i.e., qubits or spins). Meanwhile a scalar field at the same site ($\phi_j$) is mapped directly onto the motional state of the ions (i.e., phonons). (b) Schematic depiction of the various elementary gates needed to implement Trotterized time evolution of the Yukawa model on the spin-phonon quantum computer. The CNOT gate is implemented using a single M\o lmer-S\o rensen gate along with single-qubit rotations.
    Qubits and motional modes are depicted by solid and dashed lines, respectively.
    (c) Unitary operations that make up the Trotterized time evolution under various terms of the Yukawa Hamiltonian in Eq.~\eqref{eq:Ham}.}
    \label{fig:protocol}
\end{figure*}

\section{\label{sec:theory}Results}
Consider a one-dimensional spatial lattice with $N$ sites separated by lattice spacing $b$. The lattice-regularized Yukawa-model Hamiltonian with periodic boundary conditions (PBCs) consists of:
\begin{equation}
\label{eq:Ham}
    H=H_\text{f}+H_\text{b}+H_\text{fb}.
\end{equation}
Here and throughout, we work with the natural units $\hbar=c=1$.
The purely fermionic part of the Yukawa Hamiltonian, using a Kogut-Susskind discretization of fermions~\cite{kogut1975hamiltonian} contains the nearest-neighbor hopping and staggered mass terms: 
\begin{equation}
\begin{aligned}
    H_\text{f}=&\frac{1}{4 b} \sum_{j=0}^{N-2} \left(\sigma_j^x \sigma_{j+1}^x+\sigma_j^y\sigma_{j+1}^y\right)+\\ &\frac{\chi}{4 b} \left(\sigma_{N-1}^x \sigma_{0}^x+\sigma_{N-1}^y\sigma_{0}^y\right)+\frac{m_\psi}{2} \sum_{j=0}^{N-1}(-1)^{j+1} \sigma_j^z,
\end{aligned}
\end{equation}
where the original fermionic operators are mapped to a spin operators via a Jordan-Wigner transformation (see Methods). Here, $\sigma_j$ are Pauli matrices acting at fermion qubit $j$, and $m_\psi$ is the bare mass of the fermionic field $\psi$. The hopping term across the periodic boundary involves a factor of $\chi=(-1)^{Q+1}$, where
\begin{equation}
Q=\sum_{j=0}^{N-1} \frac{(-1)^{j+1}\mathds{1}_j+\sigma^z_j}{2}
\label{eq:charge}
\end{equation}
is the charge operator. In this form, the fermionic state $\ket{0} \equiv \ket{\uparrow}$ ($\ket{1}\equiv \ket{\downarrow}$) at even sites represents the absence (presence) of a fermion with charge $-1$, to be called an electron. Moreover, the fermionic state $\ket{0}$ ($\ket{1}$) at odd (even) sites represents the presence (absence) of an antifermion with charge $+1$, to be called a positron. The purely bosonic part of the Hamiltonian describes the free Hamiltonian of a scalar field, expressed as the sum over $N$ momentum modes. Written in terms of the bosonic creation ($a_{\mathrm{m}}^{\dagger}$) and annihilation ($a_{\mathrm{m}}$) operators, this is the well-known quantum-harmonic-oscillator Hamiltonian\footnote{Throughout, we use a math font for site and qubit indices while a roman font for the bosonic mode indices.}
\begin{equation}
    H_\text {b}=\sum_{\mathrm{m}=0}^{N-1} \varepsilon_{\mathrm{m}}\left(a_{\mathrm{m}}^{\dagger} a_{\mathrm{m}}+\frac{1}{2}\right).
\end{equation}
Here, single-particle energies are 
\begin{equation}
\varepsilon_\mathrm{m}=\sqrt{\left(\frac{2 \pi}{N b} \left(\mathrm{m}-\frac{N}{2}\right)\right)^2+m_{\phi}^2},
\end{equation}
with $m_{\phi}$ denoting the bare mass of the scalar field $\phi$. The Yukawa-interaction term takes the form
\begin{align}
H_\text{fb} = &\sqrt{\frac{g^2 b}{8 N}} \sum_{j=0}^{N-1}\left(\mathds{1}_j+\sigma_j^z\right) \sum_{\mathrm{m}=0}^{N-1} \frac{1}{\sqrt{\varepsilon_{\mathrm{m}}}} \times \nonumber\\
& \left(a_{\mathrm{m}}^{\dagger} e^{-i \frac{2 \pi (j+1)}{N}\left(\mathrm{m}-\frac{N}{2}\right)}+a_{\mathrm{m}} e^{i \frac{2 \pi (j+1)}{N}\left(\mathrm{m}-\frac{N}{2}\right)}\right),
\label{eq:H-fb}
\end{align}
where $g$ is the coupling between the fermions and scalar fields.

Elements of the Yukawa model and its mapping to the trapped-ion spin-phonon quantum simulator of this work are depicted in Fig.~\ref{fig:protocol}. The spins in the Yukawa model are encoded in two internal electronic states of the ions, which form an array of qubits. Meanwhile, the bosons map to the ions' collective normal modes of motion, whose quantized excitations correspond to phonons. For a chain of $N$ ions, there are $3N$ motional modes. We reserve $N$ modes for spin-spin interactions and a further $N$ modes for encoding the bosons of the Yukawa Hamiltonian. (A remaining set of $N$ modes are unused.)

We digitize the time evolution in the Yukawa theory using a first-order Trotter-Suzuki expansion. Within this framework, evolution under the first term of the Hamiltonian in Eq.~\eqref{eq:Ham}, i.e., $H_\text{f}$, couples spins $i$ and $j$ or rotates each spin $j$ independently. These terms can be implemented using the well-known M\o lmer-S\o rensen (MS) gate ($e^{-i\theta\sigma_i^x\sigma_j^x/2}$) combined with single-qubit rotations ($e^{-i\theta\sigma_j^{x/y/z}/2}$). Evolution under the second term of the Hamiltonian in Eq.~\eqref{eq:Ham}, i.e., $H_\text{b}$, imparts phases to $a$ and $a^\dagger$. This operation can be implemented classically in the same way $e^{-i\theta \sigma_j^z/2}$ gates are implemented (see Methods). Evolution under the third term of the Hamiltonian in Eq.~\eqref{eq:Ham}, i.e., $H_\text{fb}$, couples spin $j$ and bosonic mode $\mathrm{m}$. These terms can be implemented by spin-dependent kicks with arbitrary angles and phases ($e^{-i\theta\sigma^y_j(e^{i\varphi}a_\mathrm{m}+e^{-i\varphi}a^\dagger_\mathrm{m})/2}$) combined with single-qubit gates, see Methods. Evolution under $H_\text{fb}$ also involves terms with no operations on spins (terms proportional to $\mathds{1}_j$), which is implemented using the same spin-phonon operation but with an ancilla ion `anc' prepared in the positive eigenstate of $\sigma^y_\text{anc}$~\cite{davoudi2021toward}.

\begin{figure*}[t!]
    \centering
    \includegraphics[width=1.0\textwidth]{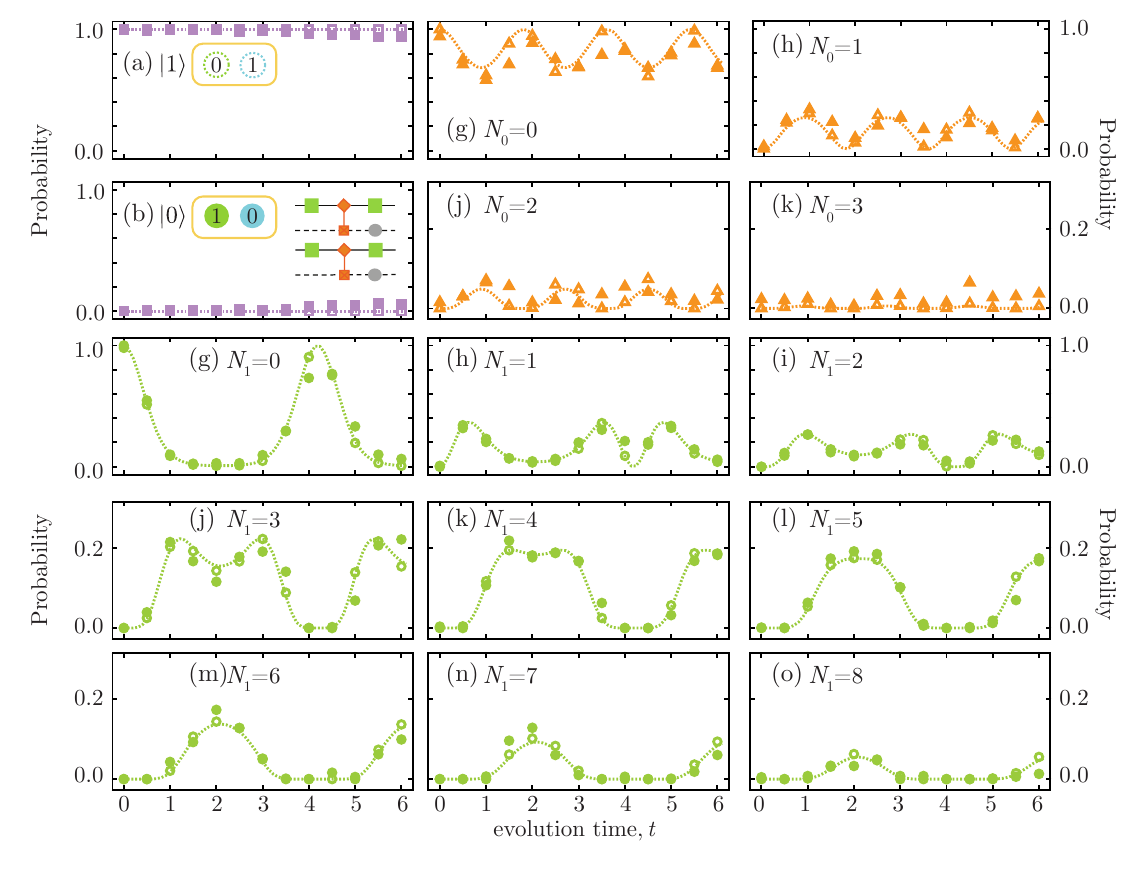} 
    \caption{\emph{Experimental results for $N=2$ and $Q=0$}. (a-b) Probability of each qubit state as a function of evolution time, as defined in Eq.~\eqref{eq:Ps-def}. Solid symbols are experimental results for Trotterized time evolution, open symbols result from noiseless emulation of the circuits, and the dashed line is the result of numerical evaluation of the continuous evolution. Both circuit emulation and continuous evolution use a cutoff of $\Lambda=15$ on each mode. The insets in (a) and (b) are schematic representations of the fermionic occupation of the lattice, prior to the mapping in Eq.~\eqref{eq:map2}, corresponding to each qubit state. The presence or absence of an electron at the left site is indicated by a solid or empty green circle while the presence or absence of a positron at the right site is indicated by a filled or open blue circle. The number inside the circle indicates the qubit state. The inset in (b) also includes a graphical depiction of the circuit for one Trotter step. (See Fig.~\ref{fig:protocol} for the symbols introduced). (c-f) Probability of mode 0 (with $\varepsilon_0\approx 3.48$) having occupation $N_0$ as defined in Eq.~\eqref{eq:Pm-def}. Note the different vertical axis ranges. (g-o) Probability of mode 1 (with $\varepsilon_1=1.5$) hosting $N_1$ bosons as defined in Eq.~\eqref{eq:Pm-def}. Uncertainties for both spin and phonon measurements were estimated using a bootstrap procedure, resulting in error bars smaller than the symbol size.}
    \label{fig:N2}
\end{figure*}

\subsection{Fermion-boson dynamics of an $N=2$ system
\label{sec:N2}}
The Yukawa Hamiltonian commutes with the charge operator in Eq.~\eqref{eq:charge}. To simplify hardware implementation, one can restrict the dynamics to a particular charge sector, such that the corresponding states can be mapped to fewer qubits. For $N=2$, the $Q=0$ sector contains two spin states: either no electron or positron, $\ket{01}$, or one of each, $\ket{10}$. These states can be encoded in a single qubit via the mapping
\begin{equation}
\begin{aligned}
&|10\rangle \mapsto |0\rangle,
\\
&|01\rangle \mapsto |1\rangle.
\label{eq:map2}
\end{aligned}
\end{equation}
Within this sector, as shown in Methods, the Hamiltonian reduces to
\begin{align} 
H^{N=2}_{Q=0}=& m_\psi \sigma^z +\varepsilon_0 a_0^{\dagger} a_0+\varepsilon_1 a_1^{\dagger} a_1
\nonumber\\
&\sqrt{\frac{g^2 b}{4}} \left(\frac{1}{\sqrt{\varepsilon_0}} a_0^{\dagger} \sigma^z+\frac{1}{\sqrt{\varepsilon_1}} a_1^{\dagger}+\text{h.c.}\right),
\label{eq:H-N2}
\end{align}
where the Pauli operator with no index acts on the single qubit. Mode 1 is uncoupled from the qubit, and the spin only undergoes phase evolution.

To simulate evolution under this Hamiltonian, we implement the Trotter circuit shown in the inset of Fig.~\ref{fig:N2}(b) for a number of Trotter steps. The circuit acts on one system qubit, one ancilla qubit, and two motional modes, with each qubit separately interacting with a given motional mode. We study the Trotterized evolution starting from the vacuum state $\ket{\Psi(0)} = \ket{s=0}\otimes\ket{N_0=0}\otimes\ket{N_1=0}$. Here, the kets from left to right are the state of the single system qubit, the occupation-number state of mode 0, and of mode 1. 
Since this state is not an eigenstate of the full Hamiltonian in Eq.~\eqref{eq:Ham}, a finite energy is imparted into the system, yielding nontrivial quench dynamics.

We track the post-quench dynamics by measuring the spin and each of the motional modes independently. Concretely, we measure the probability that the state at time $t$, $\ket{\Psi(t)}=e^{-iHt}\ket{\Psi(0)}$, projects onto one of the two spin basis states $\ket{s}$ with $s=0,1$ regardless of its boson occupation:
\begin{align}
P_{s}(t) \coloneq \sum_{N_0,N_1}\left|\langle s| \otimes \langle N_{\mathrm{0}}|\otimes \langle N_{\mathrm{1}}|\Psi(t)\rangle\right|^2,
\label{eq:Ps-def}
\end{align}
where $|N_0\rangle$ and $|N_1\rangle$ are the Fock basis states for mode 0 and mode 1, respectively. Similarly, we measure the probability that the time-evolved state turns into a state with motional-mode occupation $N_\mathrm{m}$ in mode $\mathrm{m}$ regardless of the occupation of the other mode, and of the spin state:
\begin{align}
P_{N_{\mathrm{m}}}(t) \coloneq \sum_s \sum_{N_{\mathrm{m}' \neq \mathrm{m}}} \left|\langle s | \otimes \langle N_{\mathrm{m}'}|\otimes \langle N_{\mathrm{m}}|\Psi(t)\rangle\right|^2.
\label{eq:Pm-def}
\end{align}

We quantum-compute the above probabilities for the following parameter set in the Yukawa model: $b=1, m_\psi=1, m_\phi=1.5$, and $g=4$. The corresponding free-boson energies are $\varepsilon_0 \approx 3.48$ and $\varepsilon_1 = 1.5$. The system is evolved up to $T=6$ with twelve Trotter steps of size $\delta t =0.5$. Times are in lattice-spacing ($b$) unit, which is set to one. For the spin measurements, each data point consists of $1000$ experimental shots. A fitting procedure was performed to obtain data points for the motional-mode measurements, see Methods.

The experimental results are displayed in Fig.~\ref{fig:N2}, along with the numerically evaluated probabilities for the Trotterized and continuous evolutions (using a boson cutoff of $\Lambda=15$). The experimental spin and boson dynamics agree well with theoretical expectations. Remarkably, the phonon-occupation probabilities for each mode are rather accurately reproduced by experiment, even for occupation numbers as high as eight. Such a high occupation indicates that a large cutoff $\Lambda > 8$ in this mode would have been required if a qubit-based encoding had been adopted, requiring at least four qubits to binary-encode this mode alone. A higher occupation in mode $\mathrm{m}=1$ is consistent with the fact that this mode has a lower mode frequency $\varepsilon_\mathrm{m}$, plus it couples more strongly to the fermions since the Yukawa coupling is proportional to $\varepsilon_{\mathrm{m}}^{-1/2}$ according to Eq.~\eqref{eq:H-fb}.

\begin{figure*}[t!]
    \centering
    \includegraphics[width=1.0\textwidth]{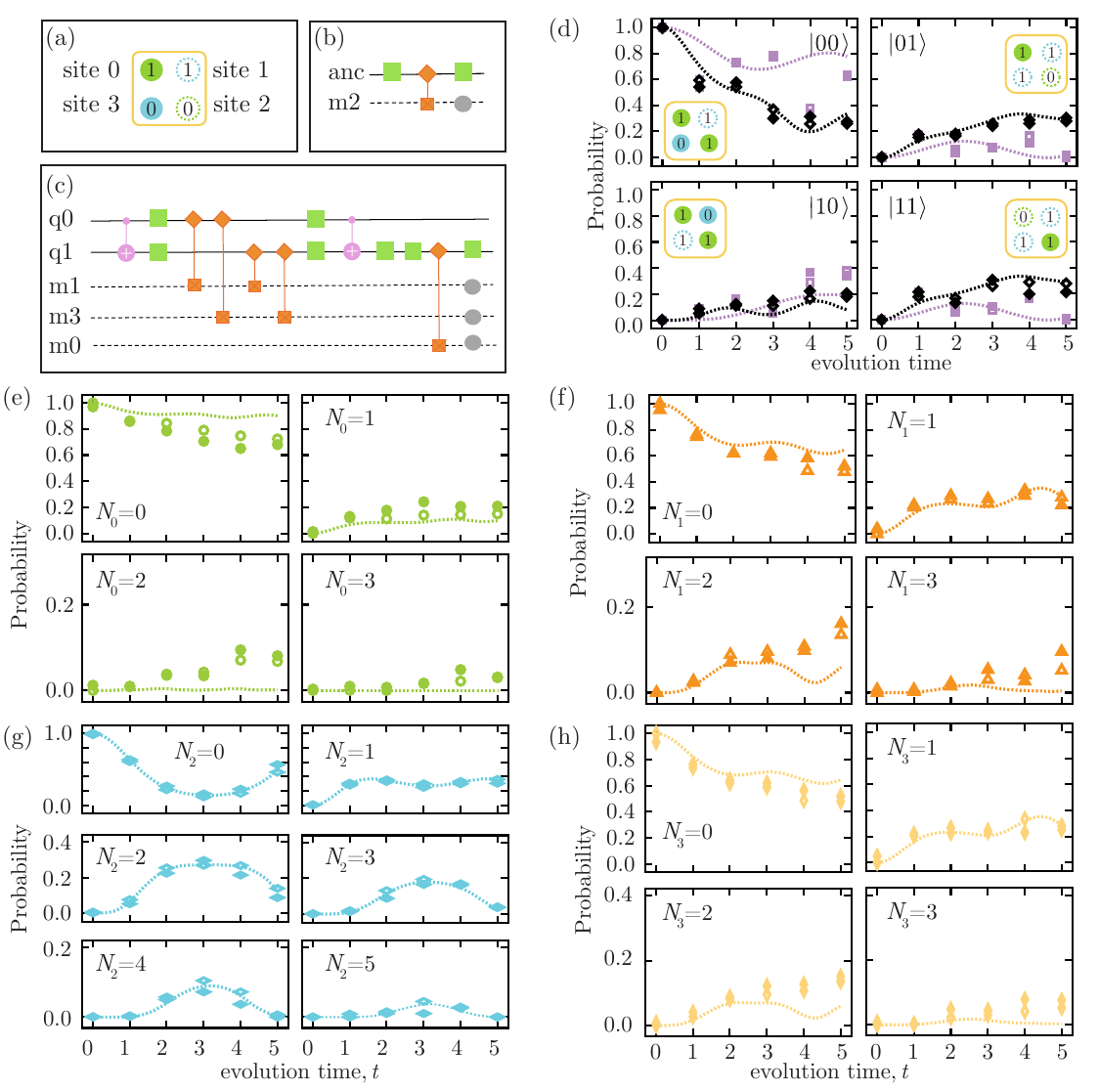} 
    \caption{\emph{Experimental results for $N=4$ and $Q=-1$}. (a) The simulated lattice with four sites prior to the mapping in Eq.~\eqref{eq:map4}. At even sites, the presence or absence of an electron is indicated by a filled or empty green circle, respectively. Analogously, filled or empty blue circles at odd sites represent the presence or absence of a positron. The number in the circle indicates the qubit state. (b-c) The quantum circuits for one Trotter step of time evolution.  The mapping in Eq.~\eqref{eq:H-N4} allows for simulation using two separate circuits. (d) Probability of each qubit basis state as a function of evolution time as defined in Eq.~\eqref{eq:Pm-def} for $g=2$ (squares) and $g=0$ (rhombuses). Solid symbols represent experimental data for Trotterized time evolution, open symbols are the result of noiseless circuit emulations, and dashed lines result from numerical evaluation of the continuous time evolution. Circuit emulation uses a cutoff of $\Lambda=15$ on each mode while continuous evolution uses a cutoff of $\Lambda=8$. Insets show the lattice occupation corresponding to each qubit basis state. (e-h) Probability of a given mode containing a number of bosons, $N_\mathrm{m}$, as defined in Eq.~\eqref{eq:Pm-def}, with $g=2$. (e-circles) $\varepsilon_0 \approx 3.30$, (f-triangles) $\varepsilon_1 \approx 1.86$, (g-horizontal diamonds) $\varepsilon_2=1$, and (h-vertical diamonds), $\varepsilon_3 \approx 1.86$. Uncertainties for both spin and phonon measurements were estimated using a bootstrap procedure, resulting in error bars smaller than the symbol size.
    }
    \label{fig:N4}
\end{figure*}

\subsection{Fermion-boson dynamics of an $N=4$ system
\label{sec:N4}}
For $N=4$, we focus on the $Q=-1$ sector, which has four spin states: two states with one electron ($\ket{1101}$ and $\ket{0111}$) and two states with two electrons and one positron ($\ket{1011}$ and $\ket{1110}$). These states can be mapped to the computational basis states of two qubits:
\begin{equation}
\begin{aligned}
|1110\rangle &\mapsto |00\rangle, \\
|1101\rangle &\mapsto |01\rangle, \\
|1011\rangle &\mapsto |10\rangle, \\
|0111\rangle &\mapsto |11\rangle.
\label{eq:map4}
\end{aligned}
\end{equation}
In this sector, as shown in Methods, the Hamiltonian can be written as
\begin{align}
H^{N=4}_{Q=-1} &= \frac{1}{2b} \left( \sigma_1^x + \sigma_0^x \sigma_1^x \right) + m_\psi \, \sigma_1^z \nonumber\\
&+ \sqrt{\frac{g^2 b}{32}} \Bigg[
\frac{2}{\sqrt{\varepsilon_0}} \, a_0^\dagger \sigma_1^z
+ \frac{1-i}{\sqrt{\varepsilon_1}} \, a_1^\dagger \sigma_0^z
+ \frac{1+i}{\sqrt{\varepsilon_1}} \, a_1^\dagger \sigma_0^z \sigma_1^z \nonumber\\
& + \frac{2}{\sqrt{\varepsilon_2}} \, a_2^\dagger
+ \frac{1+i}{\sqrt{\varepsilon_3}} \, a_3^\dagger \sigma_0^z
+ \frac{1-i}{\sqrt{\varepsilon_3}} \, a_3^\dagger \sigma_0^z \sigma_1^z
+ \text{h.c.} \Bigg] \nonumber\\
& + \sum_{{\mathrm{m}}=0}^3 \varepsilon_\mathrm{m} a_{\mathrm{m}}^\dagger a_{\mathrm{m}}.
\label{eq:H-N4}
\end{align}
Here, mode 2 is uncoupled from the rest of the system. Thus, we simulate evolution under this Hamiltonian in two separate circuits as shown in Fig.~\ref{fig:N4}(b-c). One circuit consists of two qubits and three motional modes, and the other circuit consists of one ancilla qubit and one motional mode. Evolution under the reduced Hamiltonian in Eq.~\eqref{eq:H-N4} can be Trotterized and implemented using the gate set shown in Fig.~\ref{fig:protocol}(b), see Methods.

We quantum-compute probabilities of spin state $\ket{s}$, $P_s$, and the boson-occupation state $\ket{N_{\mathrm{m}}}$, $P_{N_\mathrm{m}}$ starting from the state $\ket{\psi(0)}=\ket{s=00}\bigotimes_{\mathrm{m}=0,\cdots,3}\ket{N_\mathrm{m}=0}$, where the probabilities are straightforward generalizations of those defined in Eqs.~\eqref{eq:Ps-def} and \eqref{eq:Pm-def}. The model parameters are set to $b=1, m_\psi=1$, $m_\phi=1$, and $g \in \{0,2\}$. The corresponding free-boson energies are $\varepsilon_0 \approx 3.30$, $\varepsilon_1\approx1.86$, $\varepsilon_2=1$, and $\varepsilon_3\approx1.86$. The system is evolved up to $T=5$ with five Trotter steps of size $\delta t = 1$.

The experimentally obtained probabilities are plotted in Fig.~\ref{fig:N4} along with the numerically evaluated quantities for Trotter ($\Lambda = 15$) and continuous ($\Lambda=8$) evolutions. Both spin and boson dynamics are captured well by the experiment. The effect of boson dynamics on spins is clearly evident upon comparing results for $g=0$ and $g=2$ results, as shown in Fig.~\ref{fig:N4}(d). The probability of occupying higher excitations increases as a function of time, as shown in Figs.~\ref{fig:N4}(e-h). The energy deposited into the system due to the quench is expected to be approximately conserved throughout the Trotterized time evolution. Part of this energy is converted to exciting more bosons. The mode that sees the strongest population change is mode 2, as it has the lowest mode frequency.

\section{\label{sec:discussion} Discussion}
This work demonstrates the advantages of a hybrid analog-digital trapped-ion quantum simulator for scaling up bosonic quantum-field-theory simulations. First, the use of motional modes to encode the bosonic degrees of freedom expands the capability of already-existing systems by leveraging available quantum resources. Assuming a binary encoding of the bosons onto qubits and a cutoff $\Lambda$ on the boson occupation, the hybrid approach reduces the qubit count of simulating the Yukawa model consisting of $N$ fermionic sites from $\mathcal{O}(N+N\log\Lambda)$ to $\mathcal{O}(N)$~\cite{davoudi2021toward}. It also circumvents truncation errors when mapping infinite-dimensional bosons onto finite-dimensional qubits. The gate count is also greatly reduced: implementing the fermion--scalar-field interaction term in a fully digital model requires $\mathcal{O}(N^2(\log\Lambda)^2)$ entangling gates, but a hybrid approach  requires $\mathcal{O}(N^2)$ spin-phonon gates~\cite{davoudi2021toward}, some of which can be parallelized by grouping terms that commute.

Even for the small system sizes simulated in the experiments of this work, the Hilbert-space size is significant, making classical computation on local computing clusters challenging. The reduced Hilbert-space size is $2 \times (\Lambda+1)^2$ for $N=2$ and $4 \times (\Lambda+1)^4 $ for $N=4$. For the evolution times considered in this work, the maximum deviation in the mean occupations, across all modes, is found to be less than $2\%$ when comparing the results at cutoffs $\Lambda=7$ and $\Lambda=8$. $\Lambda=8$ amounts to a Hilbert-space dimension of 26,244 (roughly equivalent to the Hilbert-space size of 15 qubits). This discussion makes it clear that quantum-field-theory simulations of real-time dynamics toward the continuum and thermodynamic limits are not tractable with classical computing, and a bosonic quantum computer is an ideal platform for such simulations. 

There are, nonetheless, experimental challenges to scaling this hybrid approach to simulate larger systems.  Most pressing is the relatively short motional coherence time, which is typically one to two orders of magnitude shorter than the spin coherence time. 
In our system, the idle spin coherence time is $\approx1.5\:$s, which reduces to $\approx~400\:$ms under Raman laser control. Our motional coherence time is $\lesssim10\:$ms. 
There are two relevant timescales of motional decoherence: motional dephasing, $\tau_\text{deph}$ and motional heating, $\tau_\text{heat}$, which are approximate analogs of a qubit $\tau_2$ and $\tau_1$ times, respectively. Recent interest in using trapped-ion devices for continuous-variable quantum computing has launched concerted efforts to mitigate the technical noise which limits motional coherence, so far reaching coherence times of $\tau_\text{deph}\approx50\:$ms and $\tau_\text{heat}\approx5\:$s~\cite{valahu2024benchmarking, matsos2025universal}. Hence, it is plausible that further improvement will make the motional coherence times similar to qubit coherence times, in the hundreds of milliseconds~\cite{debnath2016demonstration}.

Another challenge is the ability to address individual motional modes. As the number of ions in a chain increases, the frequency spacing between modes decreases, requiring a weaker spin-phonon coupling to avoid driving unwanted interactions with neighboring motional modes. In turn, this increases the duration of spin-phonon gates. The extent of this problem can be reduced by developing ion traps with stronger confinement, directly increasing the motional-mode frequency separation. This problem can be further mitigated through the use of pulse shaping where the displacement of unwanted modes can be explicitly nulled~\cite{blumel2021efficient}.

Finally, strong driving of the motional modes will result in significantly more expansive spatial wavefunctions, which may complicate aspects of experimental control. First, accurate synthesis of spin-spin and spin-phonon interactions rely on the validity of the Lamb-Dicke approximation~\cite{sorensen1999quantum}. If this approximation is invalid, naive implementation of pulse synthesis will yield incorrect coherent operations. However, supporting pulses or more advanced pulse shaping can cancel these incorrect operations. Additionally, large motional excitation may cause the ion to sample a wide variation of laser beam intensities or phases, driving spin dephasing. Again, these effects can be canceled with the use of pulse shaping, such as stimulated Raman adiabatic passage~\cite{vitanov2017stimulated}, and by improved trap technology that realizes stronger radial confinement. Despite these potential challenges, coherent control over excitations as high as 100 has been achieved~\cite{mccormick2019quantum}.

The Yukawa model of this work involves only fermion-boson interactions linear in the bosonic operators. More realistic models of bosonic fields in nature, nonetheless, involve inter-boson interactions. For example, pions as well as gauge bosons self interact. In general, simulating arbitrary bosonic theories requires a set of Gaussian and non-Gaussian bosonic operations~\cite{marshall2015quantum,yeter2022quantum,briceno2024toward,gupta2025euclidean}. Such operations have been recently demonstrated on trapped-ion quantum computers~\cite{katz2023demonstration,matsos2025universal,buazuavan2024squeezing,hou2024coherent}. Control of local, rather than global, motional modes has also been shown \cite{Debnath2018} with the potential for additional control of phonon-phonon interactions~\cite{Porras2004,davoudi2021toward}. 

With the gate set successfully implemented in this work, and the extended gate set mentioned above, spin-phonon quantum computers can start to address a wide range of physical phenomena in quantum field theories and beyond. For example, by providing access to both an expansive bosonic reservoir and tunable system-reservoir interactions, such computers can enable tests of the quantum-thermodynamics framework for strongly coupled systems in and out of equilibrium~\cite{jarzynski2004nonequilibrium,seifert2016first,davoudi2025work,davoudi2024quantum}. Through accessing large system-reservoir Hilbert spaces, these simulations can also explore thermalization paradigms in quantum systems~\cite{srednicki1994chaos,khaymovich2019eigenstate,abanin2019colloquium,kaufman2016quantum,geraedts2016many,mueller2022thermalization,mueller2025quantum}. Last but not least, they may enable a deeper look into the entanglement structure of bosonic field theories~\cite{klco2021geometric,klco2023entanglement,klco2023entanglementII,klco2024identification}. These simulations may require more advanced motional-mode-measurement techniques, such as phase-sensitive measurements that give access to correlations between motional states~\cite{Chen2023}. Together with the ability to access qudit degrees of freedom~\cite{meth2023simulating}, trapped-ion spin-phonon quantum simulators pave the way toward practical quantum advantage in simulating quantum field theories.

\emph{Note:} A lattice-field-theory simulation of the $Z_2$ gauge model, also relying on a hybrid spin-phonon ion-trap architecture, is reported in a recent experiment by the Ion Trap Group at the University of Oxford~\cite{saner2025real}.

\section{\label{sec:method} Methods}
In the following, we will describe in more detail the Yukawa model; the derivation of the Hamiltonians in the selected charge sectors; as well as the hardware and our experimental sequence for implementing the Hamiltonian time evolution. 

\subsection{The (1+1)D Yukawa model on a spatial lattice}
Consider a one-dimensional spatial lattice with $N$ sites, lattice spacing $b$, and with periodic boundary conditions (PBCs) imposed. The Hamiltonian of the lattice-regularized Yukawa theory consists of
\begin{equation}
    H=H_\text{f}+H_\text{b}+H_{\text{fb}},
\end{equation}
where the purely-fermionic Hamiltonian,
\begin{equation}
   H_\text{f}=\sum_{j=0}^{N-1}\left[\frac{i}{2 b}\left(\psi_j^{\dagger} \psi_{j+1}-\psi_{j+1}^{\dagger} \psi_j\right)+m_\psi(-1)^{j+1} \psi_j^{\dagger} \psi_j\right],
\end{equation}
describes the hopping term and the mass term of one flavor of staggered fermions~\cite{kogut1975hamiltonian} with mass $m_\psi$. Note that PBCs impose the identification $\psi_{N} \equiv \psi_0$. \\\\
The free scalar field can be quantized to obtain a representation in terms of the bosonic (harmonic-oscillator) creation $(d_\mathrm{m}^{\dagger})$ and annihilation $(d_\mathrm{m})$ operators,
\begin{equation}
\phi_j=\frac{1}{\sqrt{N b}} \sum_{\mathrm{m}=-\frac{N}{2}}^{\frac{N}{2}-1} \frac{1}{\sqrt{2 \tilde\varepsilon_\mathrm{m}}}\left(d_\mathrm{m}^{\dagger} e^{-i 2 \pi \mathrm{m} (j+1) / N}+\text{h.c.}\right),
\label{eq:phi-mode-expansion}
\end{equation}
where $\mathrm{m}$ labels the corresponding momentum mode $p_\mathrm{m} = 2 \pi \mathrm{m} /(N b)$, and $\tilde\varepsilon_\mathrm{m}=\sqrt{\left(\frac{2 \pi \mathrm{m}}{N b}\right)^2+m_{\phi}^2}$ is the corresponding energy,\footnote{This is the continuum limit of the lattice dispersion relation. For simplicity, we adopt this form and refrain from working with the lattice dispersion relation.} with $m_{\phi}$ being the bare mass of the scalar field. The resulting Hamiltonian for the free scalar field,
\begin{equation}
H_\text{b}=\sum_{\mathrm{m}=-N / 2}^{N / 2-1} \tilde\varepsilon_\mathrm{m}\left(d_\mathrm{m}^{\dagger} d_\mathrm{m}+\frac{1}{2}\right),
\end{equation}
describes the energy of $N$ uncoupled quantum harmonic oscillators.

Finally, the interacting fermion scalar-field Hamiltonian is
\begin{equation}
H_{\text{fb}}=g b \sum_{j=0}^{N-1} \psi_j^{\dagger} \phi_j \psi_j,
\end{equation}
where the field $\phi_j$ can be written in terms of its Fourier-mode expansion in Eq.~\eqref{eq:phi-mode-expansion}.

To perform an analog-digital simulation of the Yukawa model on a trapped-ion system, one can map the staggered fermionic fields to spin degrees of freedom through a Jordan-Wigner transformation: $\psi_j=\prod_{l<j}(i\sigma^z_l)\sigma^-_j$ and $\psi_j^\dagger=\prod_{l<j}(-i\sigma^z_l)\sigma^+_j$ with $\sigma^\pm_j=\frac{1}{2}(\sigma^x_j\pm i \sigma^y_j)$.\footnote{With this convention for the Jordan-Winger transformation, the 1 (0) eigenvalue of $\psi_j^\dagger \psi_j$ maps to the 1 ($-1$) eigenvalue of $\sigma_j^z$, which is associated with the computational basis state $\ket{0}$ ($\ket{1}$), as used in the main text.} The $N$ bosonic modes can be mapped to a set of $N$ motional modes: $d_\mathrm{m}^{\dagger}=:a^{\dagger}_{\mathrm{m}+N/2}$, with energies $\tilde\varepsilon_\mathrm{m}=:\varepsilon_{m+N/2}$ ,for $\mathrm{m}=-N/2,-N/2+1,\cdots, N/2-1$. 

The mapped Hamiltonians read (up to immaterial constant terms):
\begin{equation}
\begin{aligned}
    H_\text{f}=&\frac{1}{4 b} \sum_{j=0}^{N-2} \left(\sigma_j^x \sigma_{j+1}^x+\sigma_j^y\sigma_{j+1}^y\right)+\\ &\frac{\chi}{4 b} \left(\sigma_{N-1}^x \sigma_{0}^x+\sigma_{N-1}^y\sigma_{0}^y\right)+\frac{m_\psi}{2} \sum_{j=0}^{N-1}(-1)^{j+1} \sigma_j^z,
    \label{eq:H-f-methods}
\end{aligned}
\end{equation}
where $\chi=(-1)^{Q+1}$, with $Q$ defined in Eq.~\eqref{eq:charge}, is a factor that accounts for the eigenvalue of the Jordan-Wigner string of $\sigma^z$ Pauli operators when the fermion hops across the boundary, 
\begin{equation}
    H_\text {b}=\sum_{\mathrm{m}=0}^{N-1} \varepsilon_{\mathrm{m}}\left(a_{\mathrm{m}}^{\dagger} a_{\mathrm{m}}+\frac{1}{2}\right),
        \label{eq:H-b-methods}
\end{equation}
and
\begin{align}
H_\text{fb} = &\sqrt{\frac{g^2 b}{8 N}} \sum_{j=0}^{N-1}\left(\mathds{1}_j+\sigma_j^z\right) \sum_{\mathrm{m}=0}^{N-1} \frac{1}{\sqrt{\varepsilon_{\mathrm{m}}}} \times \nonumber\\
& \left(a_{\mathrm{m}}^{\dagger} e^{-i \frac{2 \pi (j+1)}{N}\left(\mathrm{m}-\frac{N}{2}\right)}+a_{\mathrm{m}} e^{i \frac{2 \pi (j+1)}{N}\left(\mathrm{m}-\frac{N}{2}\right)}\right).
    \label{eq:H-fb-methods}
\end{align}

\subsection{The charge-sector Hamiltonians for $N=2,4$ \label{sec:simp}}
In the staggered fermion formulation, we adopt the convention that even sites in state $|1\rangle$ are occupied by a fermion of charge $-1$ (called an electron here), while odd sites in state $|0\rangle$ are occupied by an antifermion of charge $+1$ (called a positron here). With this convention, the staggered charge operator takes the form given in Eq.~\eqref{eq:charge}. The hopping term creates and annihilates fermion-antifermion pairs while conserving the total staggered charge. Consequently, $Q$ is preserved under the Yukawa-Hamiltonian dynamics. To simplify implementation on quantum hardware, we focus on a fixed charge subsector and map the states within this subsector onto a reduced number of qubits. Such a mapping also leads to shorter circuit depths and a reduced number of entangling gates. In the following, we show how the reduced Hamiltonians in the two cases in the main text, i.e., Eqs.~\eqref{eq:H-N2} and \eqref{eq:H-N4}, are deduced.
\begin{figure*}[t!] 
    \centering
    \includegraphics[width=0.775 \textwidth]{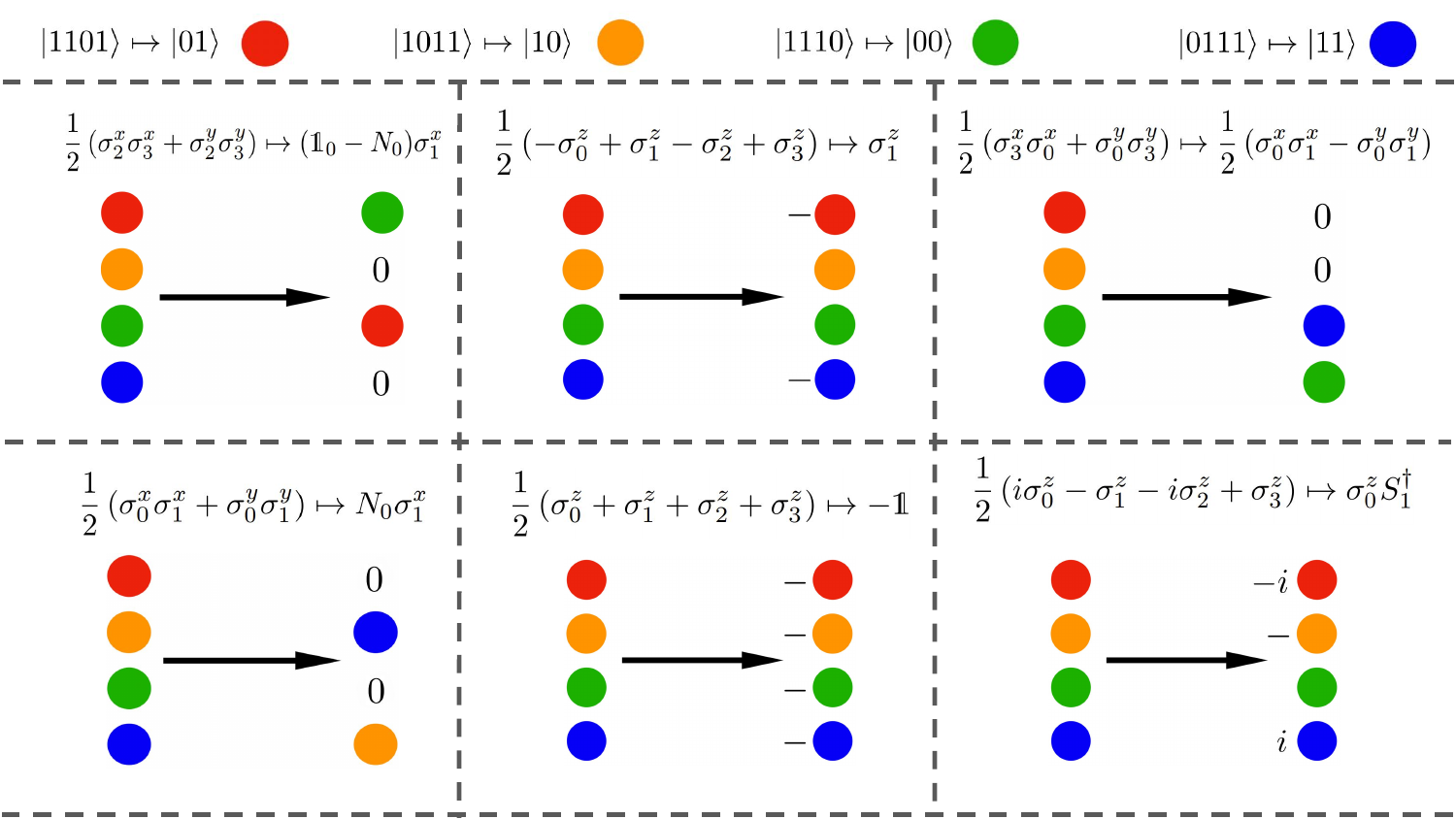} 
   \caption{\emph{The mapping of the $Q=-1$ states and operators for $N=4$.} The mapping of the four basis states in the $Q=-1$ sector to the basis states of two qubits is shown in the top. Shown in the table are action of the spin operator in the original basis and the corresponding operator in the reduced basis for each of the spin-operator combinations appearing in Eq.~\eqref{eq:H-N4-before-mapping}. Here, $N_i = \dfrac{\mathds{1}_i - \sigma^z_i}{2}$ and
 $S_i = \dfrac{1+i}{2}\,\mathds{1}_i + \dfrac{1-i}{2}\,\sigma^z_i$. `$0$' indicates that the corresponding basis state is annihilated by the action of the operator.}
    \label{fig:mapping-N4}
\end{figure*}
\subsubsection{N=2}
Using Eqs.~\eqref{eq:H-f-methods}-\eqref{eq:H-fb-methods}, the $N=2$ Yukawa Hamiltonian with PBCs is
\begin{equation}
\begin{aligned}
H_{N=2} & =\frac{m_\psi}{2}\left(-\sigma_0^z+\sigma_1^z\right) \\
& +\sqrt{\frac{g^2 b}{16}}\left[\left(-\frac{1}{\sqrt{\varepsilon_0}} a_0^{\dagger}+\frac{1}{\sqrt{\varepsilon_2}} a_1^{\dagger}\right) \sigma_0^z+\text {h.c.}\right] \\
& +\sqrt{\frac{g^2 b}{16}}\left[\left(\frac{1}{\sqrt{\varepsilon_0}} a_0^{\dagger}+\frac{1}{\sqrt{\varepsilon_1}} a_1^{\dagger}\right) \sigma_1^z+\text {h.c.}\right] \\
& +\sqrt{\frac{g^2 b}{4}} \left(\frac{1}{\sqrt{\varepsilon_1}} a_1^{\dagger}+\text {h.c.}\right) \\
& +\varepsilon_0 a_0^{\dagger} a_0+\varepsilon_1 a_1^{\dagger} a_1,
\end{aligned}
\end{equation}
where we have removed the immaterial bosons' zero-point energy. Note that the hopping term vanishes due to PBCs. We restrict ourselves to the $Q=0$ subsector with two states $|01\rangle$ and $|10\rangle$, which are mapped to a single qubit via  $|01\rangle \mapsto |1\rangle$ and $|10\rangle \mapsto |0\rangle$. Within this subsector, the Hamiltonian terms can be projected onto the corresponding operations on a single qubit. For example, the mass term becomes
\begin{equation}\label{eq:block-tikz}
\frac{m_\psi}{2}\big(-\sigma_0^z + \sigma_1^z\big)
=
m_\psi\,
\begin{tikzpicture}[baseline={(m.center)}]
  \matrix (m) [matrix of math nodes,left delimiter={[},right delimiter={]},ampersand replacement=\&]{
    0 \& 0 \& 0 \& 0 \\
    0 \& -1 \& 0 \& 0 \\
    0 \& 0 \& 1 \& 0 \\
    0 \& 0 \& 0 \& 0 \\
  };
  \node[draw,dotted,rounded corners,inner sep=1pt,fit=(m-2-2)(m-3-3)] {};
\end{tikzpicture}.
\end{equation}
Here, the marked 2×2 block corresponds to the $Q=0$ subspace, which maps directly to the single-qubit Pauli-Z operator. After applying this projection to all terms, the resulting Hamiltonian in the $Q=0$ subsector takes the form
\begin{align} 
H^{N=2}_{Q=0}=& m_\psi \sigma^z +\varepsilon_0 a_0^{\dagger} a_0+\varepsilon_1 a_1^{\dagger} a_1
\nonumber\\
&\sqrt{\frac{g^2 b}{4}} \left(\frac{1}{\sqrt{\varepsilon_0}} a_0^{\dagger} \sigma^z+\frac{1}{\sqrt{\varepsilon_1}} a_1^{\dagger}+\text{h.c.}\right),
\label{eq:H-N2-Methods}
\end{align}
which is the form in Eq.~\eqref{eq:H-N2}.

\subsubsection{N=4}
Using Eqs.~\eqref{eq:H-f-methods}-\eqref{eq:H-fb-methods}, the $N=4$ Yukawa Hamiltonian with PBCs is 
\begin{align}
H & = \frac{1}{4 b}\Big(
  \sigma_0^x \sigma_1^x + \sigma_0^y \sigma_1^y +
  \sigma_1^x \sigma_2^x + \sigma_1^y \sigma_2^y +
  \sigma_2^x \sigma_3^x + \sigma_2^y \sigma_3^y  \nonumber\\[1mm]
&\qquad + \sigma_3^x \sigma_0^x + \sigma_3^y \sigma_0^y
\Big) + \frac{m_\psi}{2}\big(-\sigma_0^z + \sigma_1^z - \sigma_2^z + \sigma_3^z\big) \nonumber\\[1mm]
& + \sum_{\mathrm{m}=0}^{3} \varepsilon_\mathrm{m}  a_\mathrm{m}^\dagger a_\mathrm{m}  + \sqrt{\frac{g^2 b}{32}} \times \nonumber\\[1mm]
& \Bigg\{ \bigg[(\mathds{1}_0 + \sigma_0^z)
\Big(-\frac{a_0^\dagger}{\sqrt{\varepsilon_0}} + \frac{ia_1^\dagger}{\sqrt{\varepsilon_1}} + \frac{a_2^\dagger}{\sqrt{\varepsilon_2}} - \frac{ia_3^\dagger}{\sqrt{\varepsilon_3}} \Big) 
\nonumber\\[1mm]
& + (\mathds{1}_1 + \sigma_1^z)
\Big( \frac{a_0^\dagger}{\sqrt{\varepsilon_0}} - \frac{a_1^\dagger}{\sqrt{\varepsilon_1}} + \frac{a_2^\dagger}{\sqrt{\varepsilon_2}} - \frac{a_3^\dagger}{\sqrt{\varepsilon_3}} \Big) 
\nonumber\\[1mm]
& + (\mathds{1}_2 + \sigma_2^z)
\Big(-\frac{a_0^\dagger}{\sqrt{\varepsilon_0}} - \frac{ia_1^\dagger}{\sqrt{\varepsilon_1}} + \frac{a_2^\dagger}{\sqrt{\varepsilon_2}} + \frac{ia_3^\dagger}{\sqrt{\varepsilon_3}} \Big)
\nonumber\\[1mm]
& + (\mathds{1}_3 + \sigma_3^z)
\Big( \frac{a_0^\dagger}{\sqrt{\varepsilon_0}} + \frac{a_1^\dagger}{\sqrt{\varepsilon_1}} + \frac{a_2^\dagger}{\sqrt{\varepsilon_2}} + \frac{a_3^\dagger}{\sqrt{\varepsilon_3}} \Big)\bigg] 
+\text{h.c.} \Bigg\},
\end{align}
where again the bosons' zero-point energy is removed. Rearranging the terms yields:
\begin{align}
H & = \frac{1}{4 b}\Big(
  \sigma_0^x \sigma_1^x + \sigma_0^y \sigma_1^y +
  \sigma_1^x \sigma_2^x + \sigma_1^y \sigma_2^y +
  \sigma_2^x \sigma_3^x + \sigma_2^y \sigma_3^y  \nonumber\\[1mm]
&\qquad + \sigma_3^x \sigma_0^x + \sigma_3^y \sigma_0^y
\Big)
+ \frac{m_\psi}{2}\big(-\sigma_0^z + \sigma_1^z - \sigma_2^z + \sigma_3^z\big) \nonumber\\[2mm]
& \quad \; + \sqrt{\frac{g^2 b}{32}}\Bigg[ \bigg( \frac{4}{\sqrt{\varepsilon_3}}\,a_3^\dagger 
+ \frac{-\sigma_0^z + \sigma_1^z - \sigma_2^z + \sigma_3^z}{\sqrt{\varepsilon_0}}\, a_0^\dagger \nonumber\\[1mm]
&\quad \; +
\frac{i\sigma_0^z - \sigma_1^z - i\sigma_2^z + \sigma_3^z}{\sqrt{\varepsilon_1}}\, a_1^\dagger +
\frac{\sigma_0^z + \sigma_1^z + \sigma_2^z + \sigma_3^z}{\sqrt{\varepsilon_2}}\, a_2^\dagger \nonumber\\[1mm]
&\quad \; +
\frac{-i\sigma_0^z - \sigma_1^z + i\sigma_2^z + \sigma_3^z}{\sqrt{\varepsilon_3}}\, a_3^\dagger \bigg)
+ \text{h.c.} \Bigg] \nonumber\\[2mm]
& + \sum_{\mathrm{m}=0}^{3} \varepsilon_\mathrm{m}  a_\mathrm{m}^\dagger a_\mathrm{m} .
\label{eq:H-N4-before-mapping}
\end{align}
The $Q=-1$ sector has four states $\ket{1110}$, $\ket{1101}$, $\ket{1011}$, and $\ket{0111}$, which can be mapped to the four basis states of two qubits $\ket{00}$, $\ket{01}$, $\ket{10}$, and $\ket{11}$, respectively. The various combinations of spin operators in the Hamiltonian in Eq.~\eqref{eq:H-N4-before-mapping} can be mapped to their corresponding two-qubit operators in this charge subsector, as shown in Fig.~\ref{fig:mapping-N4}. The resulting Hamiltonian is
\begin{align}
H^{N=4}_{Q=-1} &= \frac{1}{2b} \left( \sigma_1^x + \sigma_0^x \sigma_1^x \right) + m_\psi \, \sigma_1^z \nonumber\\
&+ \sqrt{\frac{g^2 b}{32}} \Bigg[
\frac{2}{\sqrt{\varepsilon_0}} \, a_0^\dagger \sigma_1^z
+ \frac{1-i}{\sqrt{\varepsilon_1}} \, a_1^\dagger \sigma_0^z
+ \frac{1+i}{\sqrt{\varepsilon_1}} \, a_1^\dagger \sigma_0^z \sigma_1^z \nonumber\\
& + \frac{2}{\sqrt{\varepsilon_2}} \, a_2^\dagger
+ \frac{1+i}{\sqrt{\varepsilon_3}}\, a_3^\dagger \sigma_0^z
+ \frac{1-i}{\sqrt{\varepsilon_3}} \, a_3^\dagger \sigma_0^z \sigma_1^z
+ \text{h.c.} \Bigg] \nonumber\\
& + \sum_{{\mathrm{m}}=0}^3 \varepsilon_\mathrm{m}a_{\mathrm{m}}^\dagger a_{\mathrm{m}},
\label{eq:H-N4-Methods}
\end{align}
which is the form in Eq.~\eqref{eq:H-N4}.

\subsection{Trapped-ion system as a qubit-boson quantum simulator
\label{sec:simulator}}
The experimental system is based on a chain of $^{171}$Yb$^+$ ions in a linear Paul trap.  The qubit is encoded within the hyperfine clock transition in the $^2S_{1/2}$ manifold, with $\ket{0} \equiv \ket{F=0,m_F=0}$ and $\ket{1} \equiv \ket{F=1,m_F=0}.$  In addition to these pseudo-spins, there exist bosonic motional modes that arise from a combination of the trap potential and Coulomb repulsion among the ions, giving rise to two sets of $N$ transverse modes and one set of $N$ axial modes for a chain of $N$ ions.
\begin{figure*}[t!] 
    \includegraphics[width=1.0\linewidth]{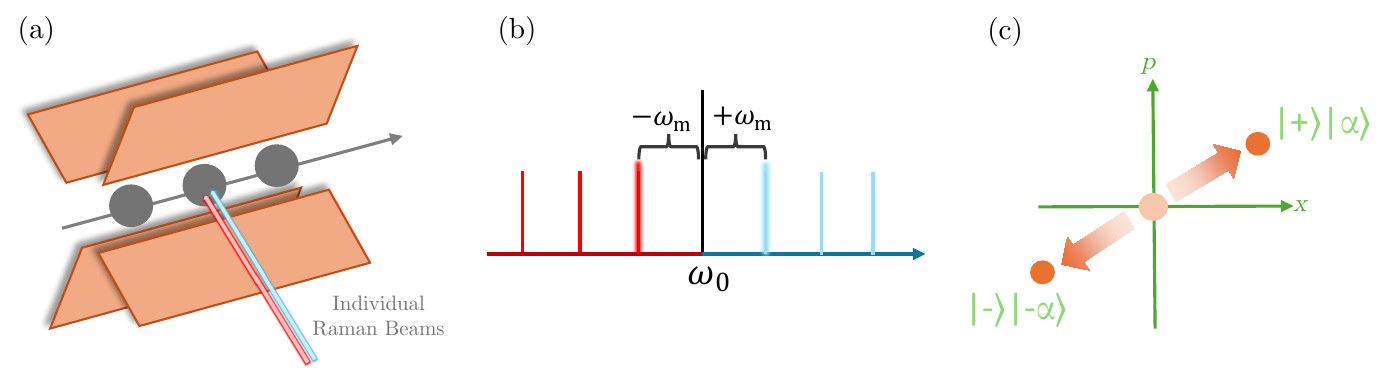} 
    \caption{\emph{Experimental implementation of spin-phonon gates.}  (a) A chain of three ions held in a blade trap. Coherent operations are performed using a Raman transition with two counter-propagating beams.  One beam is split and focused onto separate ions, allowing for phase, frequency, and amplitude control of the pulses sent to individual qubits.  (b) To implement the term for the fermion-boson coupling, a simultaneous red and blue resonant sideband operation is applied to each ion. $\omega_\mathrm{m}$ denotes the $\mathrm{m}$th mode frequency and $\omega_0$ is the qubit frequency. (c) This operation on an initial state $\ket{0}\ket{0}$ results in an entangled state $\frac{1}{\sqrt{2}}(\ket{-}\ket{-\alpha} + \ket{+}\ket{\alpha})$, where the first ket is the spin state and the second is the motional coherent state with displacement parameters $\pm \alpha$.}
    \label{fig:exp}
\end{figure*}

Coherent manipulations of the system's quantum state are achieved with off-resonant Raman transitions using two counter-propagating beams from a $355\:$nm pulsed laser~\cite{hayes2010entanglement}.  One beam addresses all ions, while the other beam is split into several beams, each focused on an individual ion after going through a multi-channel acoustic-optical modulator, allowing for individual phase, frequency, and amplitude control~\cite{debnath2016demonstration}. Single-qubit gates $R^{x/y}_j(\theta) \coloneq e^{-i\theta \sigma_j^{x/y}/2}$ on ion $j$ are performed by resonantly driving the qubit transition.  Two-qubit gates entangling ions $i$ and $j$, $\text{MS}_{i,j}(\theta) \coloneq e^{-i\theta\sigma_i^x\sigma_j^x/2}$, are performed using a M\o lmer-S\o rensen interaction \cite{molmer1999multiparticle} with frequency and amplitude-modulated pulses based on the scheme discussed in Ref.~\cite{blumel2021efficient}.  Spin-phonon interactions between ion $j$ and motional mode $\mathrm{m}$ (Fig. \ref{fig:exp}) are achieved by simultaneously driving a blue and red sideband on resonance, giving the ability to implement the gate~\cite{davoudi2021toward} 
\begin{equation} 
      \text{SNP}_{j,\mathrm{m}}(\theta,\varphi) \coloneq e^{-i\theta\sigma_j^y(e^{i\varphi}a_{\mathrm{m}}+e^{-i\varphi}a_{\mathrm{m}}^\dagger)/2}.
  \end{equation}
We use one set of $N$ transverse modes for entangling gates, and the other set for spin-phonon interactions. Details on gate calibration can be found in the Supplementary Information. 

The $e^{-i\theta\sigma^z/2}$ and $e^{-i\theta a_\mathrm{m}^\dagger a_\mathrm{m}}$ operators are implemented classically as phase shifts on individual beams.  For example, note the following identity involving the spin-phonon interaction
  \begin{equation}
      \text{SNP}(\theta,\varphi_1-\varphi_2) = e^{i\varphi_2a^\dagger a}e^{-i\theta\sigma^y(e^{-i\varphi_1}a^\dagger+e^{i\varphi_1}a)/2}e^{-i\varphi_2a^\dagger a},
  \end{equation}
where the ion and mode indices are omitted for brevity.
To implement the $e^{-i\theta a^\dagger a}$ operator, one can simply shift the $\varphi$ angle of subsequent spin-phonon gates by adjusting the laser phase.  This results in a leftover $e^{-i\varphi' a^\dagger a}$ term at the end, which is diagonal in the Fock basis and does not affect the measurement result.

\subsection{Circuit decomposition and experimental sequence
\label{sec:expt}}
We implement the Hamiltonians in Eq.~\eqref{eq:H-N2} and Eq.~\eqref{eq:H-N4} using first-order Trotterization and choose a term ordering that minimizes the number of entangling gates after circuit simplification. For $N=2$, the term ordering adopted is 
\begin{align}
\{\sigma^z,\sigma^z(a_0+a^\dagger_0),\sigma^z(a_1+a^\dagger_1),a_0^\dagger a_0,a^\dagger_1a_1\},
\end{align}

while for $N=4$, terms are ordered as 
\begin{align}
    \{&\sigma_1^x,\sigma_0^x\sigma_1^x,(1-i)a^\dagger_1\sigma_0^z+\text{h.c.},(1+i)a_1^\dagger\sigma_0^z\sigma_1^z+\text{h.c.},\nonumber\\
&(1+i)a^\dagger_3\sigma_0^z + \text{h.c.}, (1-i)a^\dagger_3\sigma_0^z\sigma_1^z + \text{h.c.}, \sigma_1^z,\nonumber\\
&a_0^\dagger \sigma_1^z + \text{h.c.},a^\dagger_0 a_0,a^\dagger_1 a_1,a^\dagger_3 a_3,a_2+a^\dagger_2,a^\dagger_2a_2\}.
\label{eq:N4order}
\end{align}

In order to facilitate circuit compression for $N=4$, we cast all entangling operations in terms of CNOT gates first, and then transpile into native gates after simplification. The evolution under the terms proportional to $\sigma^z_i\sigma^z_ja^\dagger_{\mathrm{m}} + \text{h.c.}$ are implemented by using the decomposition
\begin{equation}
    e^{-i\theta\sigma^z_i\sigma^z_j(a_\mathrm{m} + a^\dagger_\mathrm{m})/2}=\text{CNOT}_{i,j}e^{-i\theta \sigma_j^z(a_\mathrm{m}^\dagger+a_\mathrm{m})/2}\text{CNOT}_{i,j},
\end{equation}
where $i$ and $j$ are the control and target qubits of the CNOT gate, receptively. Similarly, evolution under the  $\sigma_i^x\sigma_j^x$ term in the $N=4$ case can be decomposed as
\begin{equation}
     e^{-i\theta \sigma_i^x\sigma_j^x/2}=\text{CNOT}_{i,j}e^{-i\theta\sigma_i^x/2}\text{CNOT}_{i,j}.
\end{equation}
The ordering chosen in Eq.~\eqref{eq:N4order} allows for the cancellation of all but two CNOT gates per Trotter step.

To further reduce the number of gates, we omit any gate that has no effect on the measurement result.  Specifically, we remove CNOT gates at the beginning of the circuit since they have no effect on the initial $\ket{00}$ state. Additionally, when measuring the spin, we remove any interactions involving $\sigma^z$ at the end of the circuit since they are diagonal in the measurement basis. For more detailed circuit specifications, see the Supplemental Information.

To reduce the effects of motional decoherence, we choose a mode mapping that minimizes the time it takes to run one Trotter step. For $N=2$, we encode the bosons onto one set of transverse modes, with mode $\mathrm{m}=0$ as the center-of-mass mode ($3.05$ MHz), and mode $1$ as the zigzag mode ($2.98$ MHz). For $N=4$, we use the other set of transverse modes, with mode $0$ as the tilt mode ($2.85$ MHz), mode $1$ as the center-of-mass mode ($2.88$ MHz), and mode $3$ as the zigzag mode ($2.80$ MHz). Mode $2$ is simulated in a separate circuit, and uses the zigzag mode of the latter set.

At the beginning of the experimental sequence, the ions are cooled close to the motional ground state ($\bar{n} \approx 0.05$) through Doppler cooling and resolved sideband cooling, preparing the initial motional state.  The qubits are then optically pumped to the $\ket{0}$ state; if necessary, single qubit rotations are then performed to prepare the desired initial spin state. Next, coherent operations corresponding to the Trotterized evolution are performed. Entangling gates are done using the set of transverse modes not used for encoding the bosonic modes so that they do not interfere. Finally, measurement is performed on either the spins or a given phonon mode. 

\begin{figure}[t!] 
    \centering
      \includegraphics[width=0.45\textwidth]{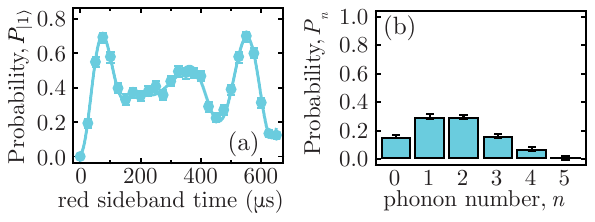} 
    \caption{\emph{Phonon probability measurement for one mode at a fixed trotter step}. Phonon measurement for $N=4$ and mode 2 at $t=3$, needed to produce the plots in Fig.~\ref{fig:N4}(g). (a) One qubit is chosen to probe the motional mode. The probability of this qubit being in $\ket{1}$ is plotted (solid circles) as a function of red sideband pulse time. The solid line is the result of fitting Eq.~\eqref{eq:phononmeas} to the experimental data. (b) The fit to Eq.~\eqref{eq:phononmeas} yields the probability that each Fock state was occupied at $t=3$. Experimental error bars are a result of bootstrapping.}
    \label{fig:phononmeas}
\end{figure}
Readout of individual ions is done through state-dependent fluorescence detection where photons from each ion are focused onto one channel of a photo-multiplier tube~\cite{olmschenk2007manipulation}.
Readout of a given motional mode $\mathrm{m}$ is performed by probing the coupling of Fock basis states to each other through a red sideband operation, driving a transition between $\ket{0}\ket{N_\mathrm{m}} \leftrightarrow \ket{1}\ket{N_\mathrm{m}-1}$, where the first ket is the spin state and the second ket is the motional Fock state. Since each mode is measured independently, here we only consider a single motional mode and denote $N_{\mathrm{m}} \equiv n$ for brevity. The Rabi frequencies associated with these transitions, driven by a single laser tone, depends on the phonon occupation number. Hence, observing the multi-frequency evolution of the spin probability allows one to extract the phonon occupation probabilities~\cite{meekhof1996generation}. This measurement process begins at the end of the Trotterized evolution. First, all spin states are pumped to $\ket{0}$, and one spin is chosen to act as the probe. This results in the state $\ket{0}\sum_{n}A_{n}\ket{n}$ where $P_{n}=|A_{n}|^2$ is the probability of projecting to the $n$th Fock state.  A resonant red sideband operation is then applied, resulting in the signal~\cite{meekhof1996generation}
\begin{equation}
    \label{eq:phononmeas}
    P_{\ket{1}}(t) = \sum_{n=1}^{\infty}P_n\sin^2\left(\frac{\Omega_nt}{2}\right)e^{-\gamma_nt},
\end{equation}
where $\Omega_n$ and $\gamma_n$ are the red sideband Rabi frequency and decay constant, respectively, for the transition involving $\ket{0}\ket{n}$. A separate experiment determines $\Omega_1$ and $\gamma_1$ by first preparing the $\ket{1}\ket{0}$ state and then driving the aforementioned red sideband transition. Such an experiment can be repeated for arbitrary Fock states. Empirically, we observe that $\Omega_n=\sqrt{n}\Omega_1$ and $\gamma_n=\gamma_1$ for the $n$ values measured in this work. 
In practice, the sum in Eq.~\eqref{eq:phononmeas} can be taken to a maximum occupation $n_\text{max}$ beyond which there is no appreciable phonon occupation.
We perform a fit to Eq.~\eqref{eq:phononmeas} to determine $P_n$ from $n=1$ to a cutoff $n_\text{max}$ and let $P_0 = 1 - \sum_{n=1}^{n_{\text{max}}}P_n$. $n_{\text{max}}$ is determined by classically simulating the expected occupations of the mode of interest in the Yukawa model, though this cutoff can be determined experimentally too.  Figure~\ref{fig:phononmeas} shows an example phonon measurement taken for $N=4$ and mode 2 after three Trotter steps ($t=3$).

\smallskip
\begin{acknowledgments}
This material is based upon work supported by the U.S. Department of Energy (DoE), Office of Science, National Quantum Information Science Research Centers, Quantum Systems Accelerator Award, no. DE-FOA-0002253 (N.M.L., A.T.T., N.H.N., X.L., A.M.G.). 
We acknowledge support from the DoE, Office of Science, Early Career Award, no. DE-SC0020271 (Z.D., S.V.K.) and no. DE-SC0024504 (N.M.L.). We acknowledge support from the National Science Foundation’s Quantum Leap Challenge Institute on Robust Quantum Simulation, award no. OMA-2120757 (Z.D., V.V., N.M.L, and N.H.N). We further knowledge support from the DoE, Office of Science, Office of Nuclear Physics, via the program on Quantum Horizons: QIS Research and Innovation for Nuclear Science, award no. DE-SC0023710 (Z.D. and V.V.). Z.D. further acknowledges support from the DoE, Office of Science, Office of Advanced Scientific Computing Research (ASCR), program in Accelerated Research in Quantum Computing, Fundamental Algorithmic Research toward Quantum Utility (FARQu). Z.D., S.V.K., and V.V. are grateful to the Department of Physics, Maryland Center for Fundamental Physics, and College of Computer, Mathematical, and Natural Sciences at the University of Maryland, College Park for their support.
S.V.K. further acknowledges support by the DoE, Office of Science, Office of Nuclear Physics, InQubator for Quantum Simulation (IQuS) (award no. DE-SC0020970), and by the DoE QuantISED program through the theory consortium ``Intersections of QIS and Theoretical Particle Physics'' at Fermilab (Fermilab subcontract no. 666484).
S.V.K. would also like to thank the Department of Physics and the College of Arts and Sciences at the University of Washington for their support.
Additional support is acknowledged from the National Science Foundation Software-Tailored Architecture for Quantum Co-Design (STAQ) Award, award no. PHY-2325080 (N.M.L). We further acknowledge early support from the DoE Office of Science, Office of Nuclear Physics, via the program on Quantum Horizons: QIS Research and Innovation for Nuclear Science, award no. DE-SC0021143 (Z.D., N.M.L, N.H.N). We thank Yingyue Zhu and Matthew Tyler Diaz for experimental support. The authors acknowledge the University of Maryland supercomputing resources (http://hpcc.umd.edu) made available for computations of this work.
\end{acknowledgments}

\bibliographystyle{apsrev4-2}
\bibliography{refs}

\appendix

\end{document}